\documentstyle[aps,amstex,epsfig,fleqn,floats]{revtex}

\begin{document}

\title{Morphological Image Analysis of Quantum \\
             Motion in Billiards}

\author{J.S. Kole, K. Michielsen and H. De Raedt}

\address{Institute for Theoretical Physics and
Materials Science Centre,\\
University of Groningen, Nijenborgh 4,
NL-9747 AG Groningen, The Netherlands}

\date{\today}
\maketitle
\begin{abstract}
Morphological image analysis is applied to the time evolution of the 
probability distribution of a quantum particle moving in two and
three-dimensional billiards.
It is shown that the time-averaged
Euler characteristic of the probability density provides 
a well defined quantity to distinguish between
classically integrable and non-integrable billiards.
In three dimensions the time-averaged mean breadth
of the probability density may also be used for this purpose.
\end{abstract}

\section{Introduction}

In the theory of quantum chaos, the connection between the chaotic
behavior of classical systems and specific properties of their
quantum mechanical counterparts has received much attention. 
The most studied issues in this context are the statistical properties 
of nuclear and atomic spectra \cite{Blumberg} and of two-(recently three)
dimensional billiards \cite{Berry,McD_etal,Lopac,Primack}.
In the mid-eighties, 
a conjecture (which became known as the BGS conjecture \cite{Bohigas})
was made that the quantal energy level spacing of (ergodic) systems 
whose classical equivalent exhibit chaotic behavior 
obeys the Wigner Gaussian-Orthogonal-Ensemble statistics,
known from random matrix theory, whereas for classically 
integrable systems the Poisson distribution applies. 
Exceptions to this conjecture are known \cite{Zakr_etal,Crehan}, 
but it is generally believed that these form a set of measure zero. 

Another way to make a connection between the chaotic behavior of classical systems
and specific properties of their quantum mechanical counterparts is to look at the
eigenstates of billiards.
Eigenstates of a stadium may exhibit structure that corresponds to
the classical orbits \cite{scars1,scars2,scars3}. These structures, called
scars, have been directly related to unstable periodic orbits 
\cite{scars4,scars5,scars6}. Visual inspection of movies that show the time evolution
of the probability distribution of a quantum particle moving in
two and three-dimensional (2D and 3D) billiards of different shapes strongly suggest that these
sequences of images may contain enough information to distinguish between
classically integrable and chaotic systems\cite{WWW}.

To our knowledge the morphology of the patterns of time (in)dependent
probability distributions has not been studied.
In this paper we propose a method, based on concepts from integral geometry,
to perform a morphological image analysis (MIA) of the probability distribution.
We use MIA to characterize the solutions of the time-dependent Schr\"odinger
equation for various 2D and 3D billiards. We demonstrate that indeed, one can
define geometrical and topological descriptors that are useful to
distinguish between classically integrable and chaotic billiards.
Our motivation to analyze time-dependent instead of stationary probability
distributions is that the former usually provide information
about many eigenstates simultaneously. By following the system over a
sufficiently long period of time, one collects information on all those
eigenstates that have non-negligible overlap with the initial state.
This ``parallelism'', absent in the stationary-state approach\cite{Baowen},
reduces the computational work.

MIA amounts to the characterization
of the geometry and topology of any pattern or series of patterns by means 
of the so-called Minkowski functionals, known from integral 
geometry \cite{int1,int2,int3}. 
These functionals are related to familiar quantities: In two (three) dimensions they
correspond to the covered area, perimeter and connectivity (volume, surface area,
integral mean curvature and connectivity) of the pattern.
In integral geometry the calculation of the Minkowski functionals is relatively
straightforward and requires little computational effort.
MIA has proven to be a powerful tool in different fields where the description
of patterns is important: Statistical physics, to describe the morphology
of porous media and complex fluids; cosmology, to analyze the large scale 
distribution of matter in the universe; chemistry, to describe the morphology 
of patterns in reaction diffusion systems \cite{Mecke2}; seismology, to describe the spatial
complexity of regional seismicity realizations\cite{Makarenko}; polymer science,
to identify and quantify the morphology of mesoscale structures in block copolymers
\cite{Michielsen1,Michielsen2}.
Furthermore, in the semi-classical limit the sum of exponentials of 
eigenfrequencies of a billiard-shaped drum can be written in terms of the 
Minkowski functionals of the billiard\cite{Kac}, and the Weyl hypothesis 
states that the number of energy levels up to an energy $E$ can be written in 
a similar manner (see e.g. \cite{weyl}).

In section 2 we briefly review the theory of Minkowski functionals
and explain the technique of MIA in more detail. We then apply MIA to the motion of a
quantum particle in various two and three-dimensional billiards and show that
an appropriate quantity can be defined to distinguish classically integrable
and chaotic billiards. Finally, we study the transition to chaos for a
system whose degree of chaos is parametrized.

\section{Morphological image analysis}

At the heart of MIA lies the computation of Minkowski functionals of an
arbitrary (thresholded) image.
Therefore, in this section we will give some mathematical
background on Minkowski functionals (closely following \cite{Mecke2}) and sketch the
outline of an algorithm to compute these functionals.

We first define the Minkowski functionals for compact convex sets.
The parallel set $K_r$ of a
compact convex set $K$ at a distance $r$ is the union of
all closed balls of radius $r$, the centers of which are points in $K$.
The well-known Euclidean measures of $K_r$, such as the area and volume, can be written as
polynomials in $r$.
For example, the parallel volume of a square $S$ and a cube $C$ with edge lengths $a$
are given by
\begin{eqnarray}
V_r(S)= & \lefteqn{a^2+4ar+\pi r^2,} \\
V_r(C)= & \lefteqn{a^3+6a^2r+3\pi ar^2 +4\pi r^3/3.}
\end{eqnarray}
In general, the volume $v^{(d)}$ in $d$ dimensions of the parallel set $K_r$
at distance $r$ of $K$ is given by the Steiner formula
\cite{int1}
\begin{equation}
v^{(d)}(K_r)=\sum_{\nu=0}^{d}\left(\begin{array}{c}d\\ \nu \end{array}\right)
W_{\nu}^{(d)}(K)r^{\nu},
\end{equation}
where $W_{\nu}^{(d)}(K)$ are the
Minkowski functionals. For the first three dimensions it can be shown
that\cite{int1}
\begin{eqnarray}
d=1 : & \lefteqn{W_0^{(1)}(K)=l(K),\ W_1^{(1)}=2\chi(K),} \\
d=2 : & \lefteqn{W_0^{(2)}(K)=A(K),\ W_1^{(2)}=\frac{1}{2}U(K), 
        \ W_2^{(2)}=\pi\chi(K),} \\
d=3 : & \lefteqn{W_0^{(3)}(K)=V(K),\ W_1^{(3)}=\frac{1}{3}S(K),\ 
        W_2^{(3)}(K)=\frac{2}{3}\pi B(K),\ W_3{(3)}(K)=\frac{4\pi}{3}\chi(K),}
\end{eqnarray}
where $l(K),A(K),U(K),V(K),S(K),B(K)$ and $\chi(K)$ denote respectively the length,
area, perimeter, volume, surface area, mean breath
and Euler characteristic or connectivity number of the convex set $K$. The mean
breath $B$ is closely related to the mean curvature $H=(R_1+R_2)/2R_1R_2$
where $R_1$ and $R_2$ are the principal radii of curvature\cite{Hyde}.
The Euler characteristic $\chi$ is related to the Gaussian curvature $G=1/R_1R_2$ and is
equal to unity for a convex body\cite{Hyde}.
The mean curvature and the Gaussian curvature are two useful measures of the curvature of a surface\cite{Hyde}.
If the local surface geometry has an elliptic, Euclidean or hyperbolic shape respectively then the
Gaussian curvature is respectively positive, zero or negative\cite{Hyde}.

Up to now, we have limited the discussion to convex sets. In order to 
deal with the more general sets one encounters in characterizing images,
we consider the convex ring $\mathcal{R}$, which is the class of all
subsets $A$ which can be expressed as finite unions of compact
convex sets $K_i$
\begin{equation}
A=\bigcup_{i=1}^l K_i .
\end{equation}
The Euler characteristic $\chi$ is defined as an additive functional on
$\mathcal{R}$, so that for $A\in \mathcal {R}$
\begin{equation}
\chi(A)=\chi\left(\bigcup_{i=1}^l K_i\right) = \sum_i \chi(K_i)
-\sum_{i<j}\chi(K_i \cap K_j)+ \ldots 
+ (-1)^{l+1}\chi(K_1 \cap\ldots\cap K_l),
\end{equation}
and where for compact convex sets $K$
\begin{equation}
\chi(K)=\begin{cases} 1 & \text{if $K\neq \emptyset$} \\
                         0 & \text{if $K=\emptyset$} \end{cases}.
\end{equation}
The Euler characteristic describes $A$ in a pure topological way, i.e.
without reference to any kind of metric. It can be shown that this definition
of $\chi$ permits all Minkowski functionals on $\mathcal{R}$ to be written
in terms of $\chi$ and the kinematical density \cite{int1}.
The functional $\chi$ as defined in integral geometry is the same as the Euler
characteristic defined in algebraic topology \cite{int1}:
For $d=2$, $\chi$ equals the number of connected components minus
the number of holes and in three dimensions $\chi$ is given by the
number of connected components minus the number of tunnels plus the
number of cavities.
For example, for a solid cube $\chi=1$, for a hollow cube $\chi=2$ and for a cube
pierced by a tunnel $\chi=0$. For multiply connected structures the Euler
characteristic is negative. For complex structures it is often difficult to
identify the number of connected components, tunnels and cavities. However,
MIA directly yields $\chi$ as will be explained below.

In order to calculate the morphological properties of a particular image
we first convert the image to a black-and-white picture using a threshold.
Pixels (voxels) with intensity below the threshold are considered to be
part of the background and the others are building up the objects in the
2D (3D) picture. According to integral geometry, the morphological
properties of the various objects building up the black-and-white picture
can be completely described in terms of Minkowski functionals \cite{int1}.
In order to calculate the Minkowski functionals in an efficient way we consider
each pixel (voxel) as the union of the disjoint collection of its interior,
faces (for the 3D case only), open edges and vertices. The values of
$A$, $U$, $V$, $S$, $B$ and $\chi$ for these single open structures can easily
be calculated \cite{comp_in_phys}. By making use of the property of additivity
of the Minkowski functionals and the fact that there is no overlap between open
bodies on a lattice, we compute the Minkowski functionals of the whole image
by starting from a complete white image, by calculating the change
in the number of interiors, faces (for the 3D case only), open edges and vertices when
one pixel (voxel) is added to the 2D (3D) image, and this until all
pixels (voxels) building up the picture are added.
The Minkowski functionals can then be computed from
\begin{eqnarray}
d=2 : & \lefteqn{A=n_s,\ U=-4n_s+2n_e,\ \chi=n_s-n_e+n_v,} \\
d=3 : & \lefteqn{V=n_c,\ S=-6n_c+2n_f,\ 2B=3n_c-2n_f+n_e,\ \chi=-n_c+n_f-n_e+n_v,}
\end{eqnarray}
where $n_s$ ($n_c$) denote the number of squares (cubes) in 2D (3D), $n_f$
counts the number of faces (in 3D only), $n_e$ and $n_v$ denote
the number of edges and vertices respectively.
From (10) and (11) it follows directly that the Minkowski functionals contain
information about local 4 (8)-point correlation functions in 2D (3D).
A more detailed description of the algorithm and an example of computer
code to compute the Minkowski functionals is given in Ref.\cite{comp_in_phys}.

\section{Method}

\begin{figure}[t]
\setlength{\unitlength}{1cm}
\begin{picture}(8.5,3.85)
\put(0.5,0.35){\epsfig{file=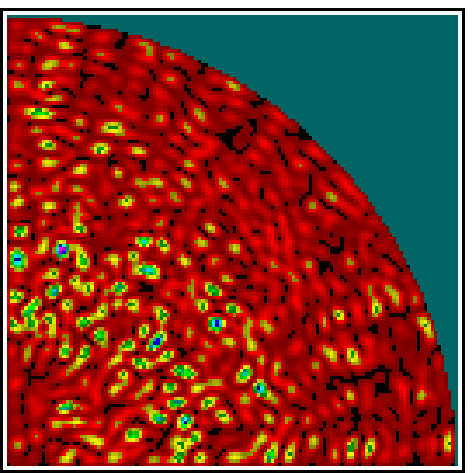,height=3.5cm,width=3.5cm}}
\put(4.5,0.35){\epsfig{file=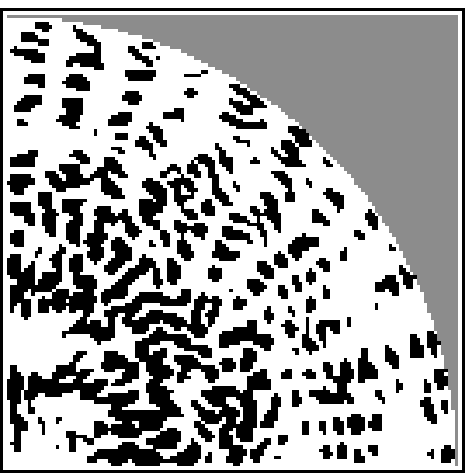,height=3.5cm,width=3.5cm}}
\end{picture}
\caption{Snapshot of
the probability density at $t=319$ (in dimensionless units, see text)
in the case of a billiard in the form of a quarter circle (left)
and its thresholded counterpart using a threshold $\theta=10\%$ (right).
The background is colored white, the objects black and the forbidden region for the
quantum particle gray.}
\label{fig:fig1}
\end{figure}

\begin{figure}[t]
\setlength{\unitlength}{1cm}
\begin{picture}(7.6,7.9)
\put(0,0.25){\epsfig{file=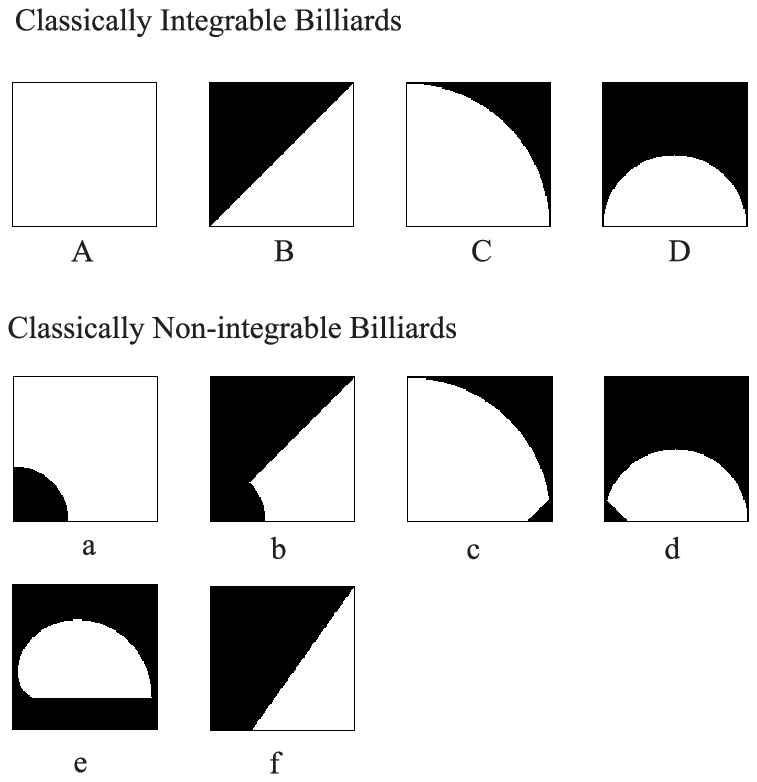,width=7.4cm,height=7.7cm}}
\end{picture}
\caption{%
The shapes of the various two-dimensional billiards studied. A: square,
B: equilateral triangle, C: quarter circle, D: semi circle,
a: perturbed square,
b: perturbed triangle, c: perturbed quarter circle, d: perturbed semi circle, 
e: semi cardiod,
f: triangle with sides which ratio is irrational.
The billiards are colored white and the forbidden regions for the quantum
particle are colored black.
In most calculations the linear size of the square was taken 
to be $13\lambda$.}
\label{fig:fig2}
\end{figure}

\begin{figure}[t]
\setlength{\unitlength}{1cm}
\begin{picture}(10,7.4)
\put(0,7.1){Classically integrable billiards}
\put(5.2,7.1){Classically Non-integrable billiards}
\put(0,4.4){\epsfig{file=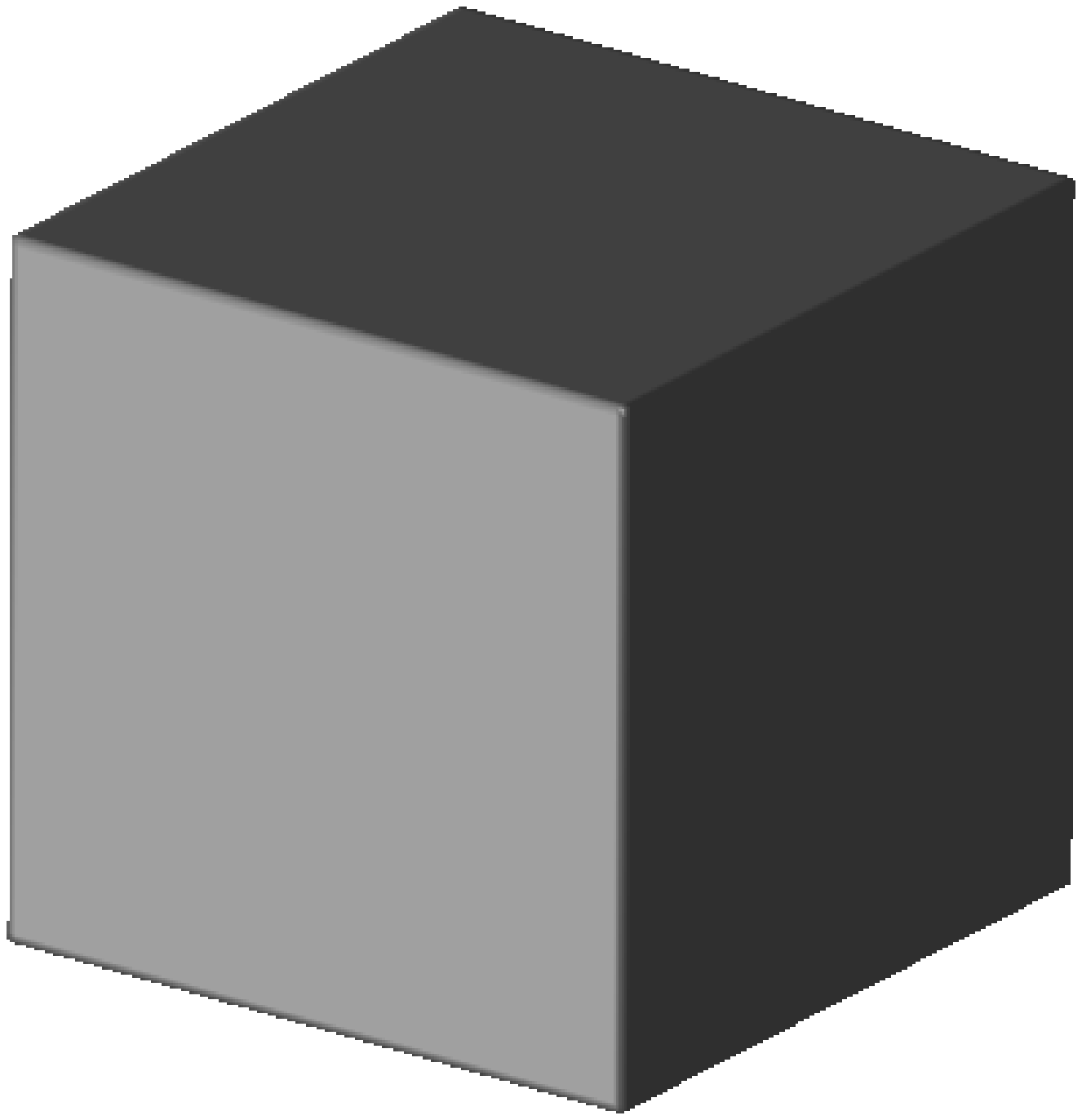,width=2.4cm,height=2.3cm}}
\put(2.7,4.4){\epsfig{file=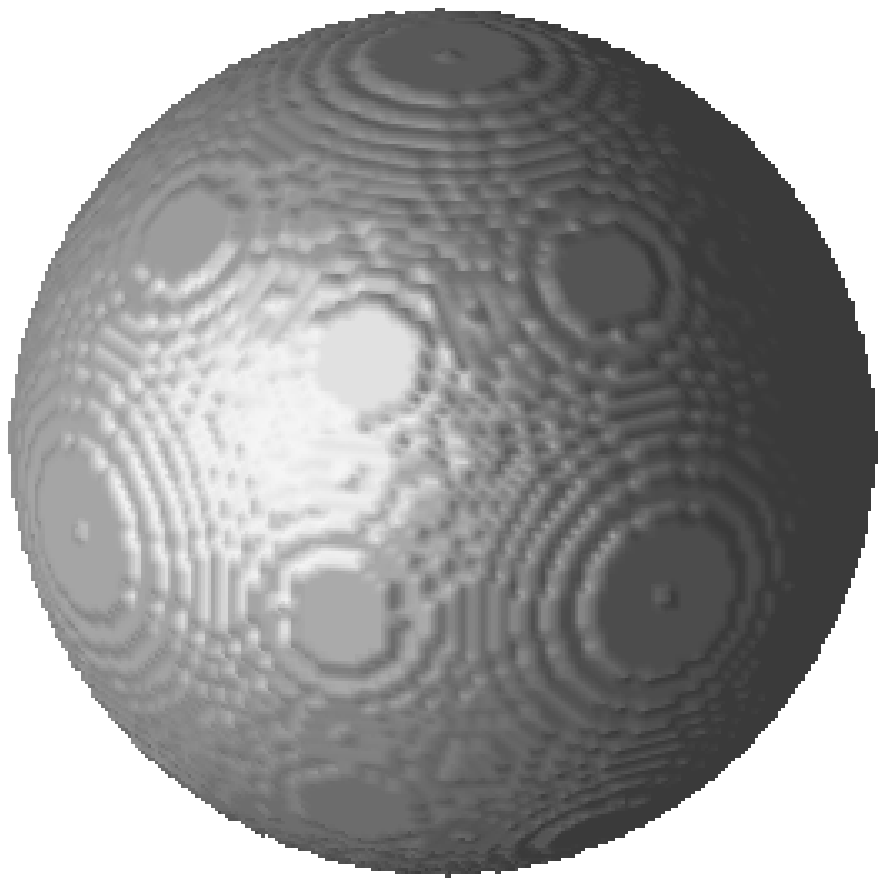,width=2cm,height=2.3cm}}
\put(5.1,4.4){\epsfig{file=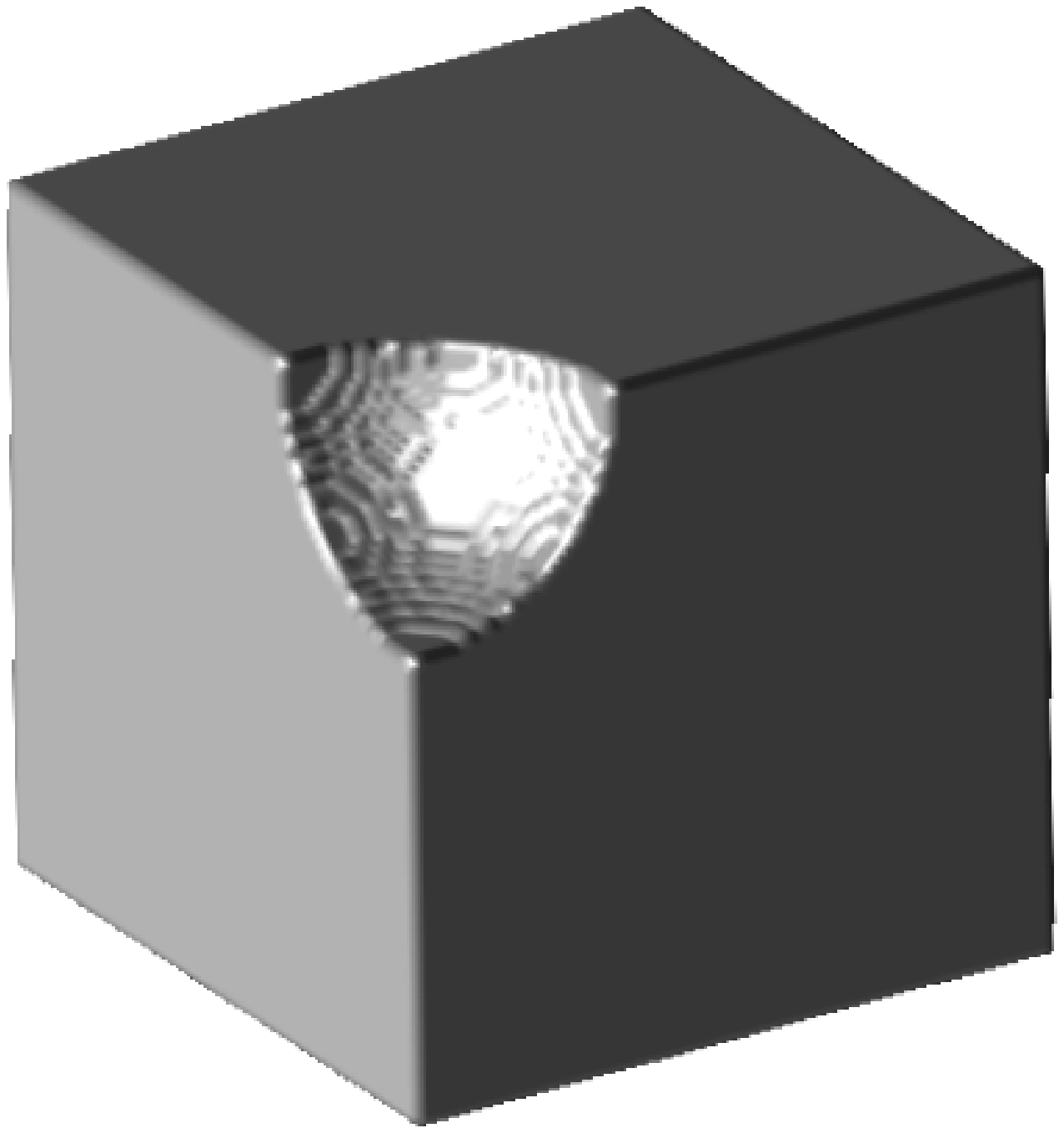,width=2.4cm,height=2.3cm}}
\put(7.8,4.4){\epsfig{file=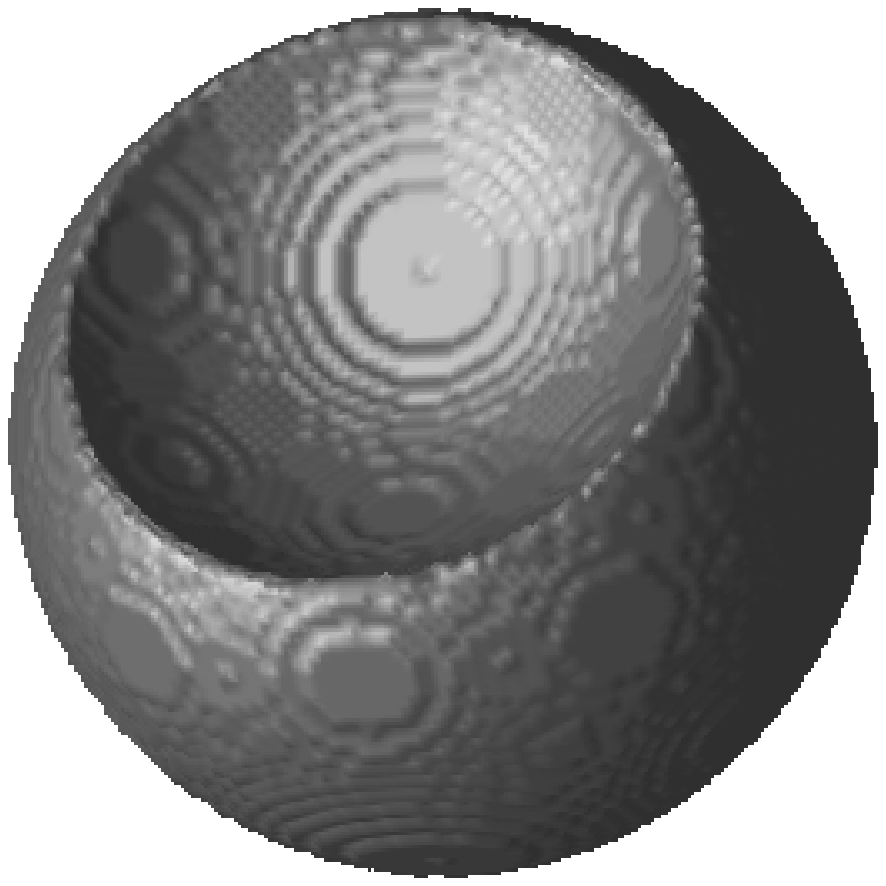,width=2.2cm,height=2.3cm}}
\put(1.1,3.9){K}
\put(3.5,3.9){L}
\put(6.3,3.9){k}
\put(8.7,3.9){l}
\put(0,0.9){\epsfig{file=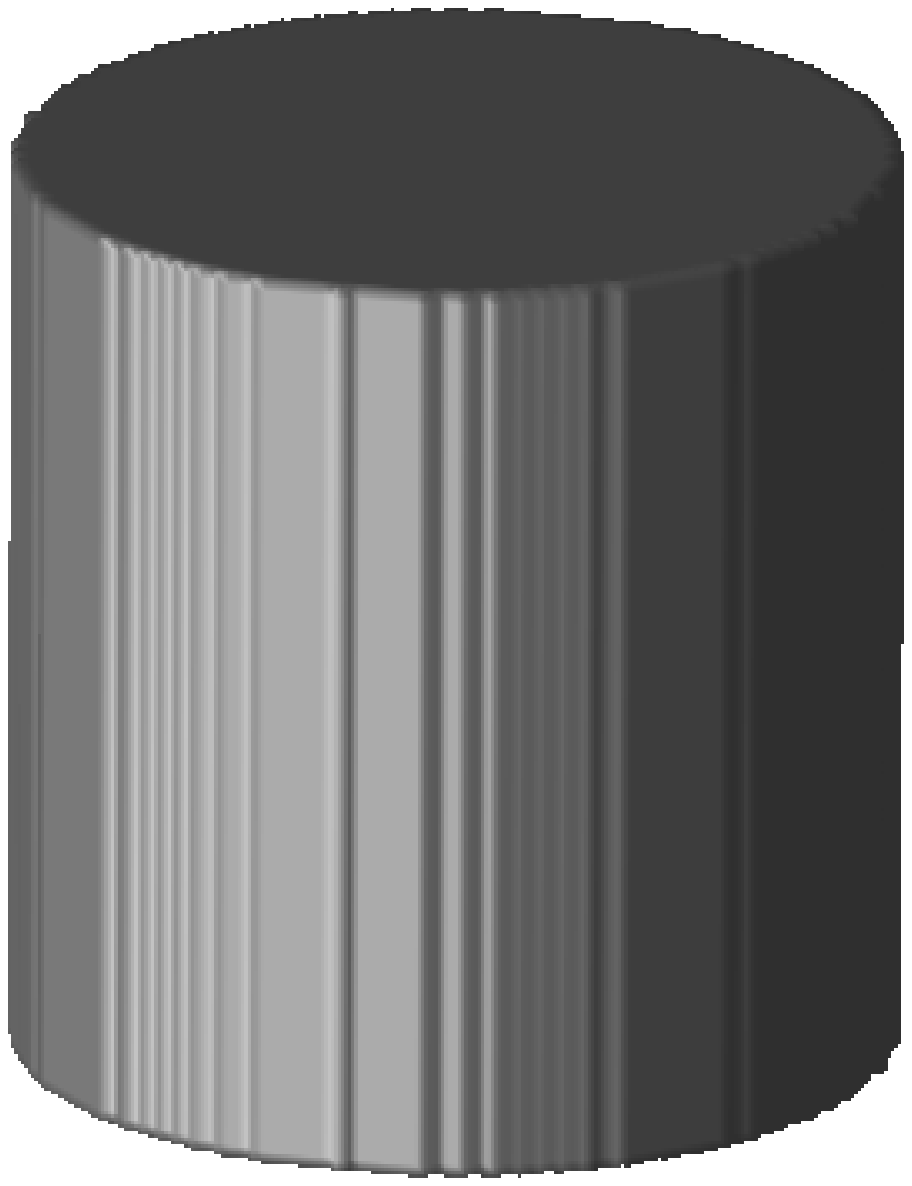,width=2.4cm,height=2.8cm}}
\put(2.7,0.9){\epsfig{file=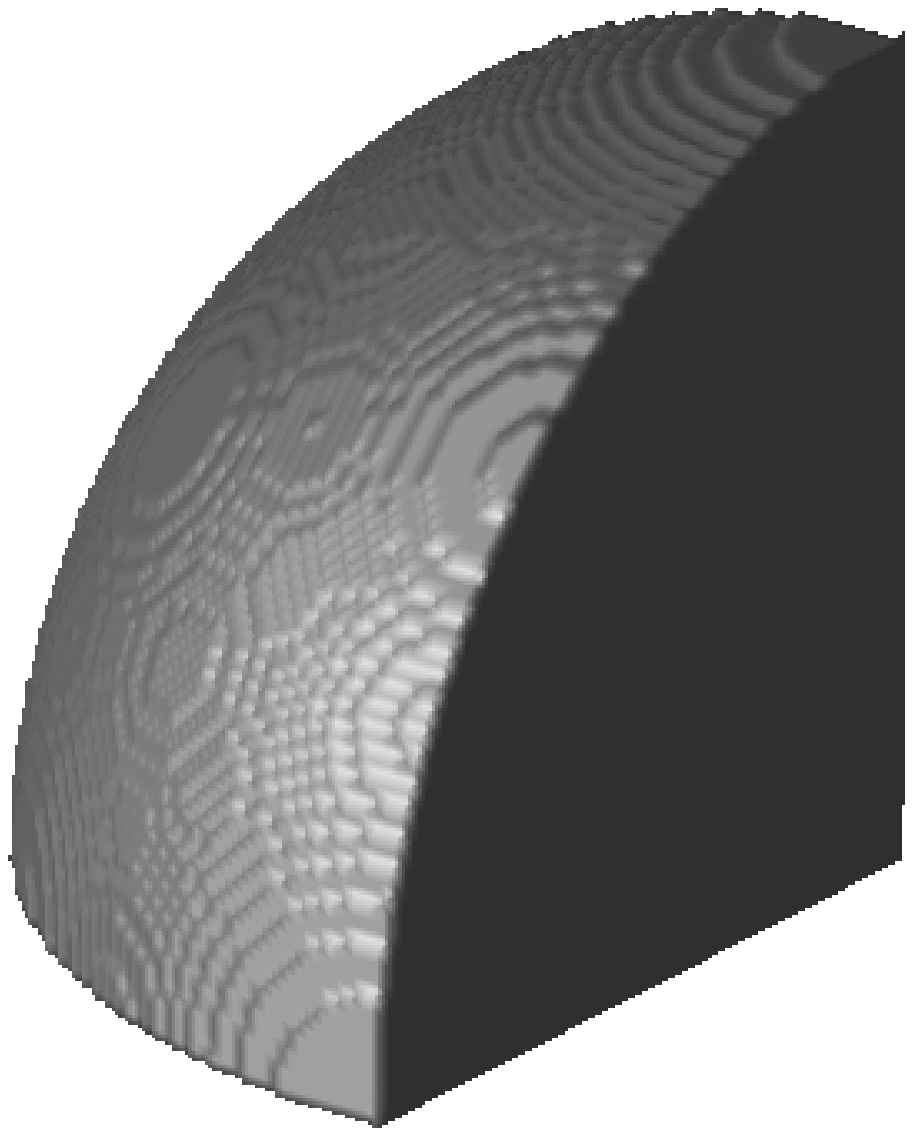,width=2.1cm,height=2.35cm}}
\put(5.1,0.9){\epsfig{file=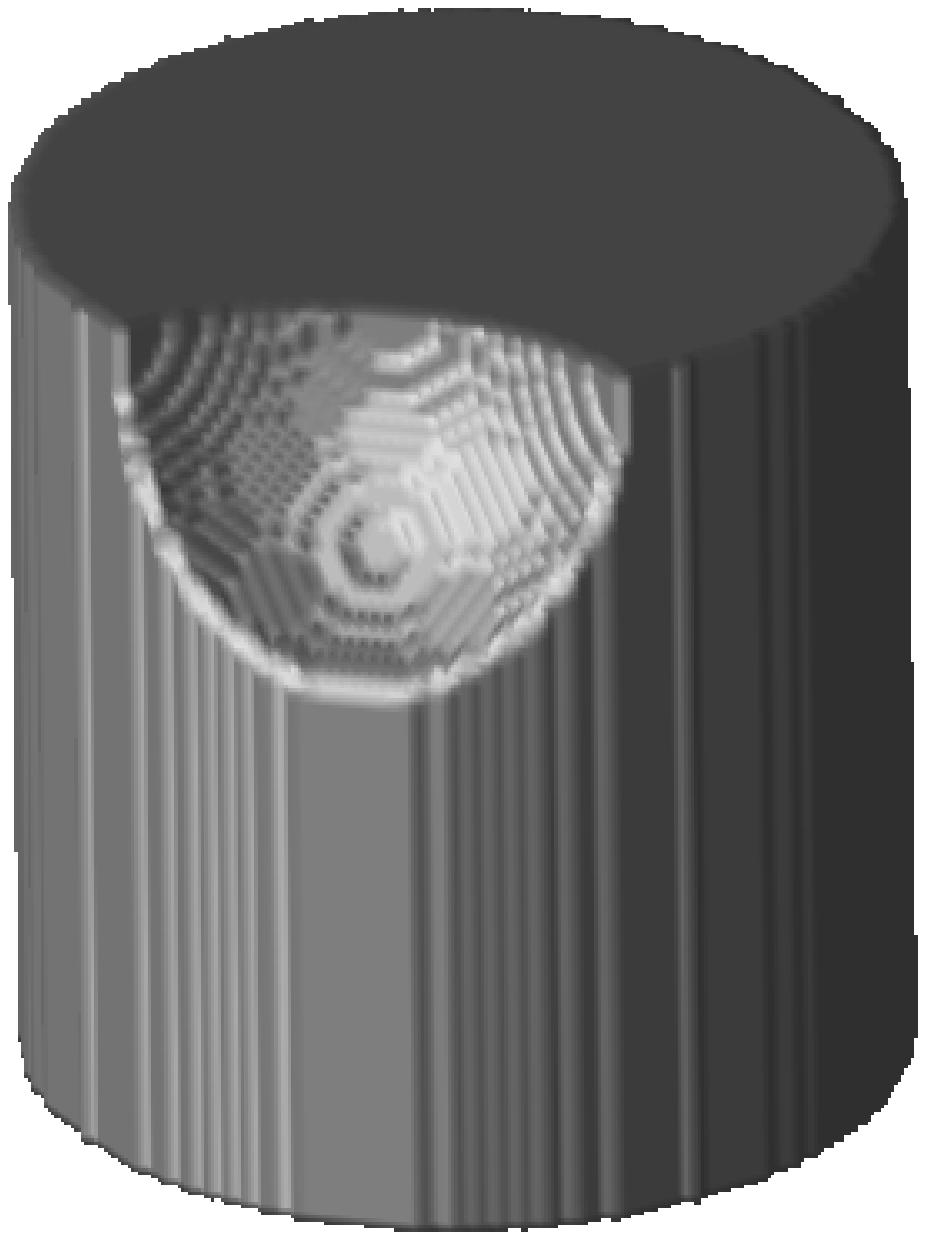,width=2.3cm,height=2.9cm}}
\put(7.8,0.9){\epsfig{file=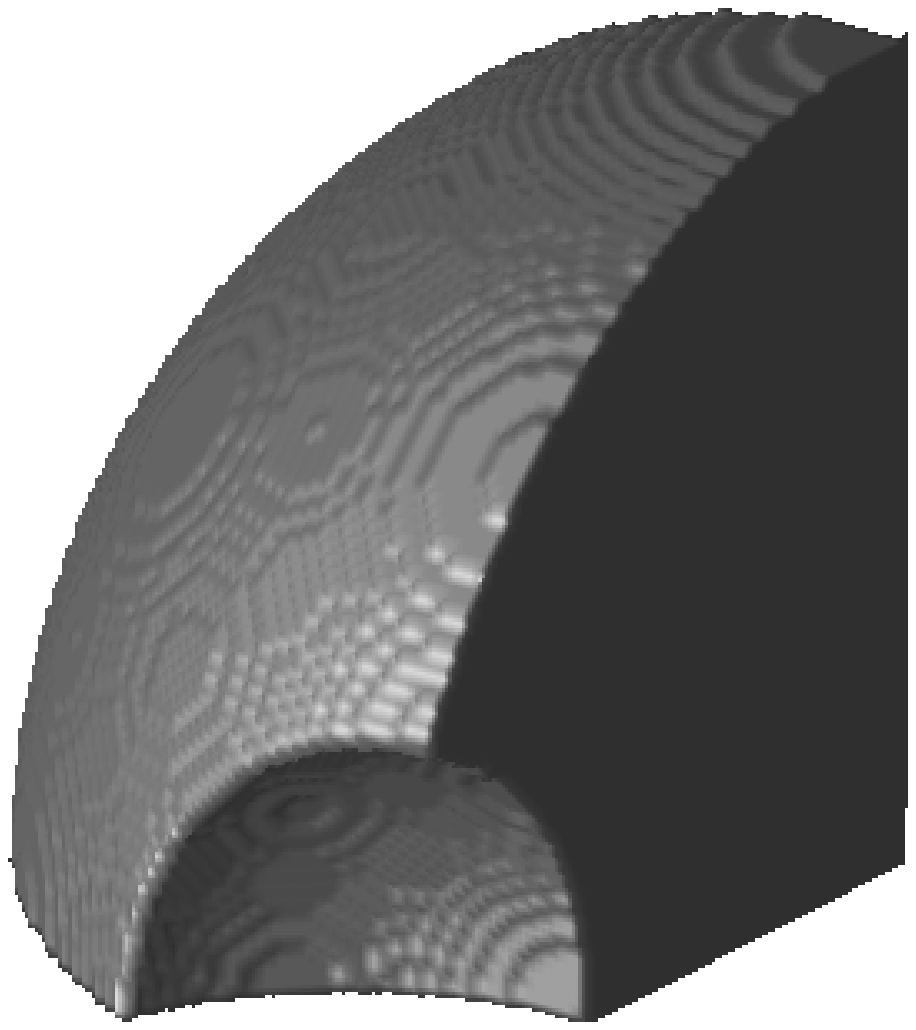,width=2.2cm,height=2.4cm}}
\put(1.1,0.4){M}
\put(3.5,0.4){N}
\put(6.3,0.4){m}
\put(8.7,0.4){n}
\end{picture}
\caption{%
The shapes of the various three-dimensional billiards studied. K: cube,
L: sphere, M: cylinder, N: one eight sphere, k: perturbed cube,
l: perturbed sphere, m: perturbed cylinder, n: perturbed one eight sphere.
In most calculations the sides of the bounding cube (not shown) were taken
to be $9.5\lambda$.}
\label{fig:pot3d}
\end{figure}

\begin{figure}[t]
\setlength{\unitlength}{1cm}
\begin{picture}(3.6,2.0)
\put(0,0.25){\epsfig{file=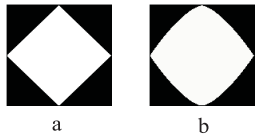,width=3.4cm,height=1.8cm}}
\end{picture}
\caption{%
Two instances of the lemon-shaped billiard, with boundaries in the $x-y$ plane defined by
$y(x)=\pm (1-|x|^{\delta})$, $x\in [-1,1]$.
A: $\delta=1$, a: $\delta=1.5$.
The billiards are colored white and the forbidden regions for the quantum
particle are colored black.
In most calculations the linear size of the square was taken
to be $13 \lambda$.}
\label{fig:fig4}
\end{figure}

The application of MIA to the time evolution of the probability distribution of a
quantum particle moving in a billiard is
straightforward. We solve the time-dependent Schr\"odinger equation for 
a particle moving in a billiard by a stable and accurate numerical 
method\cite{TDSE}. For practical purposes the results obtained are exact.
A time series of snapshots displaying the probability density can easily 
be extracted from this data. A collection of digital videos can be found 
in \cite{WWW}. Each pattern of the time series is converted into a
black-and-white picture by applying a threshold. The result of applying 
a threshold of 10\% to a
representative image of the probability density is shown in Fig.\ref{fig:fig1}.
For each black-and-white image in the time series we compute the
Minkowski functionals using the algorithm described above
and analyze the behavior of the Minkowski functionals as a function of the
control parameters such as threshold, initial conditions of the wave packet,
and shape of the billiard. 
We use Gaussian wave packets as initial states and study a variety of
classically integrable and non-integrable billiards (see Figs.\ref{fig:fig2},
\ref{fig:pot3d},\ref{fig:fig4}).
In our numerical work we express lengths in units of a fixed wavelength
$\lambda$ and rescale energy (and time, setting $\hbar=1$) such that a wave 
packet with average momentum $2\pi/\lambda$ has an average kinetic energy 
of one \cite{TDSE}.

As will become clear from the examples given below, to distinguish
between classically integrable and chaotic systems it is expedient to define the
time-averaged Minkowski functionals
\begin{equation}
X (T,\theta)= \frac{1}{\tilde{A}T}\int_{0}^{T}\chi(t,\theta)dt \quad ; \quad
{\mathbf {B}} (T,\theta)=\frac{1}{\tilde{A}T}\int_{0}^{T}B(t,\theta)dt ,
\end{equation}
and the quantities
\begin{eqnarray}
X(\theta) =& \lefteqn{\lim_{T\rightarrow\infty}X (T,\theta )\quad ; \quad
{\mathbf {B}}(\theta)=\lim_{T\rightarrow\infty}{\mathbf {B}} (T,\theta ),}\\
\mu(\chi) =& \lefteqn{\max_\theta X(\theta )\quad ; \quad
\mu(B)=\max_\theta {\mathbf {B}}(\theta ),}
\end {eqnarray}
where
$\tilde{A}=\lambda^2/A$ ($\tilde{A}=\lambda^3/V$) in two (three) dimensions.

\section{Results 2D}

In Fig. 5 we present results that illustrate the typical behavior of the Minkowski
functionals as a function of time, keeping the threshold fixed. The left panel
of Fig. 5 shows the perimeter $U(t,\theta)$ as a function of time $t$ for the
billiards (A,a) using a threshold $\theta=5\%$. Billiard A (solid line)
shows a behavior which is manifestly different from that of billiard a (dashed
line): For times $t>100$ the solid line displays large fluctuations, in contrast
to the dashed line. Other pairs of billiards (B,b) etc. show very similar, characteristic
fluctuations in the perimeter as a function of time (results not shown).
Essentially the same behavior is found for the Euler characteristic
$\chi(\theta, t)$. An example is given in the right panel of Fig.5 for the case of
billiards (B,b) and $\theta=15\%$. Hence, in conclusion, for integrable billiards $U(t,\theta)$
and $\chi (t,\theta)$ show large fluctuations while for non-integrable billiards
these large fluctuations do not occur.
\begin{figure}[t]
\setlength{\unitlength}{1cm}
\begin{picture}(17.5,4.5)
\put(0,0.2){\epsfig{file=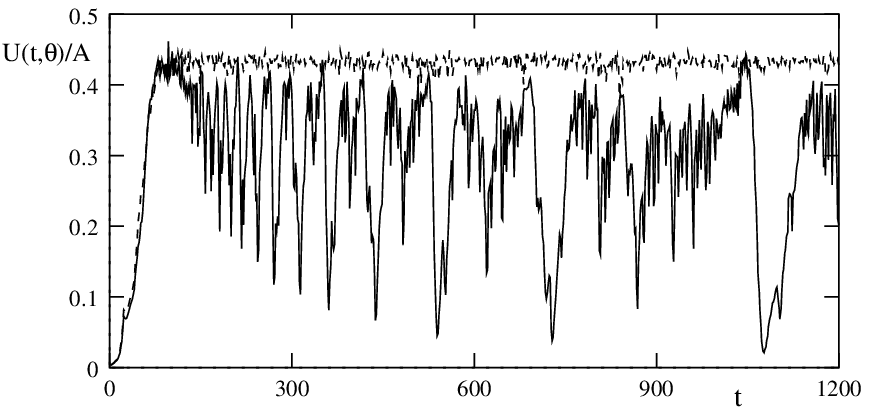,width=8.6cm,height=4cm}}
\put(8.8,0.2){\epsfig{file=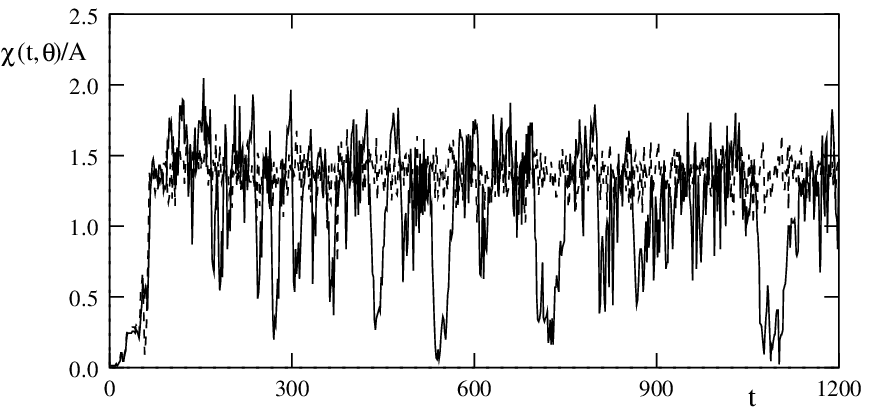,width=8.6cm,height=4cm}}
\end{picture}
\caption{%
Left: Perimeter $U(t,\theta)$ (in pixels),
normalized to the area $A$ of the billiard (in pixels), as a function of time $t$
using a threshold $\theta=5\%$.
Solid line: Billiard A. Dashed line: Billiard a.
Right: Euler characteristic $\chi(t,\theta)$,
normalized to the area of the billiard (in pixels), as a function of time $t$
using a threshold $\theta=15\%$.
Solid line: Billiard B. Dashed line: Billiard b.
Time is measured in dimensionless units (see text).}
\label{fig:fig5}
\end{figure}

\begin{figure}[t]
\setlength{\unitlength}{1cm}
\begin{picture}(8.6,4.5)
\put(0,0.2){\epsfig{file=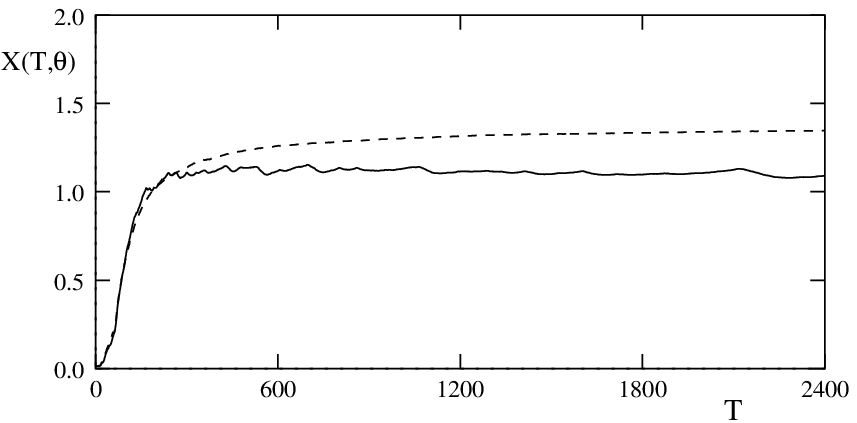,width=8.6cm,height=4cm}}
\end{picture}
\caption{%
$X(T,\theta)$ as a function of time $T$, using a threshold $\theta = 15\%$.
Solid line: Billiard B.
Dashed line: Billiard b.}
\label{fig:fig6}
\end{figure}

\begin{figure}[t]
\setlength{\unitlength}{1cm}
\begin{picture}(8.8,9)
\put(0,0.2){\epsfig{file=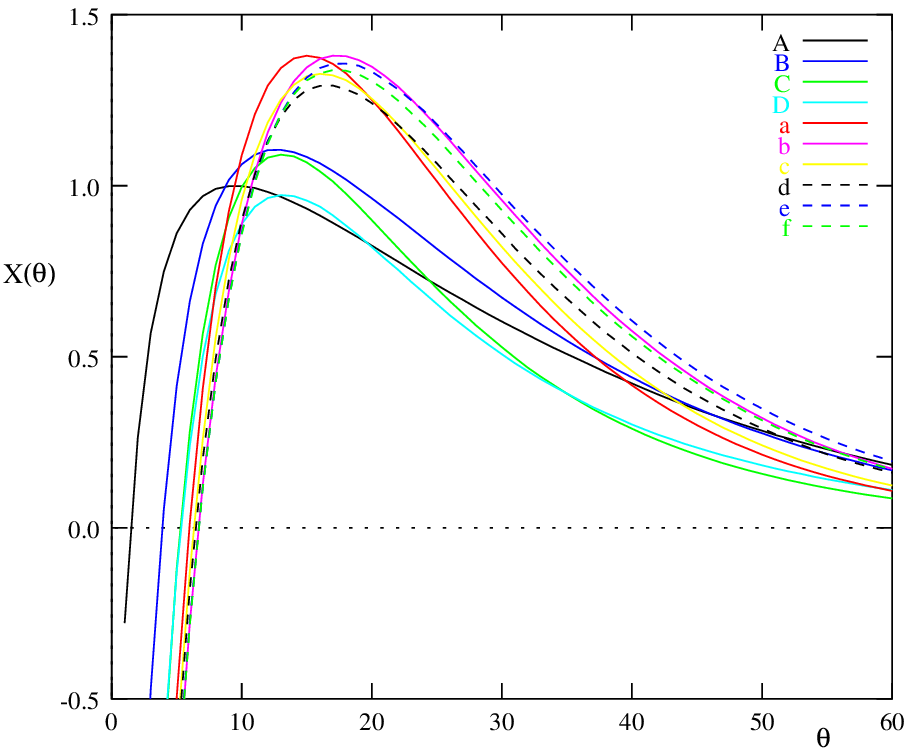,width=8.6cm,height=9cm}}
\end{picture}
\caption{%
$X(\theta)$ as function of the threshold $\theta$ (in percent), 
for all billiards shown in Fig.\ref{fig:fig2}.}
\label{fig:fig7}
\end{figure}

Since the Euler characteristic $\chi (t,\theta)$ is a global measure of the
curvatures of the objects in the black-and-white image and provides direct
quantitative information about the topology of the probability distribution
(see \cite{int2}, pp. 34, 112 and 113) we will study the behavior of
$\chi$ as a function of $t$ and $\theta$ in more detail.
For each choice of the threshold $\theta$, the cumulative time average of the
Euler characteristic $X(T,\theta)$ approaches a constant value which we denote
by $X(\theta)$. A representative example of the
behavior of $X(T,\theta)$ is shown in Fig.\ref{fig:fig6}.
In Fig.\ref{fig:fig7} we plot the results for $X(\theta)$, for each of the billiards of
Fig.\ref{fig:fig2}.
As $\theta$ approaches zero, the thresholded
black-and-white picture becomes completely filled with black pixels, hence 
$\lim_{\theta\rightarrow 0} X(\theta)=\lambda^{2}/A$, for any image. 
Obviously, for $\theta\rightarrow 0$, $X(\theta)$ does not contain any useful 
information and for reasons of clarity we therefore omit data for $\theta<1$ 
in Fig. \ref{fig:fig7}.
\begin{table}[t]
\caption{$\mu (\chi)$ and the corresponding $\theta$ for all billiards shown in Fig.\ref{fig:fig2}.}
\label{table:1}
\begin{center}
\begin{tabular}[h]{c|c|c}
 Billiard & $\theta$ & $\mu(\chi)$ \\
\hline
A & 10 & 1.00 \\
B & 13 & 1.10 \\
C & 13 & 1.09 \\
D & 13 & 0.97 \\
\hline
a & 15 & 1.38 \\
b & 17 & 1.38 \\
c & 16 & 1.33 \\
d & 17 & 1.29 \\
e & 18 & 1.36 \\
f & 17 & 1.34 \\
\end{tabular}
\end{center}
\end{table}
From Fig.\ref{fig:fig7} it is clear that the $X(\theta)$ curves
form two clusters, depending on whether the system is classically integrable 
or not. In fact, $\mu(\chi)$,
can be used to classify billiards according to their 
classical (non)integrability, as shown in Table 1.
The maxima for the classically integrable systems 
all lie in the range $[0.97,1.10]$, whereas the maxima for the classically 
non-integrable systems lie in the range $[1.29,1.38]$. 
Further analysis
indicates that the maximum of $X(\theta)$ changes linearly with the energy 
of the wavepacket (results not shown).
This scaling behavior could be trivially incorporated into the definition of 
$X(\theta)$. Calculations (results not shown) for systems up to $19\lambda
\times 19\lambda$ suggest that the system size-dependence of
$X(\theta)$ is very weak.
Therefore, we conclude that the scaling properties $\mu (\chi)$ can be employed to
test for the (non)integrability of billiards.
In contrast to what Fig.\ref{fig:fig5}(left) might suggest,
the cumulative time average and its infinite time limit of the perimeter $U$
(and also the area $A$) do not have the same properties as $X (T,\theta)$ and
$X(\theta)$ (results not shown). Hence in two dimensions only $\mu (\chi )$ can
be used to test for the (non)integrability of billiards.

\section{Results 3D}
For three dimensional billiards not much research has been done on the issue of
characterizing classical chaos on basis of some quantum mechanical property.
Very recently results of numerical work on the stability of classical
trajectories\cite{3d1}, and on 3D Sinai
billiards\cite{Primack} has been reported, but much of the field is unexplored.

\begin{figure}[b!!]
\setlength{\unitlength}{1cm}
\begin{picture}(17.4,9)
\put(0,0.2){\epsfig{file=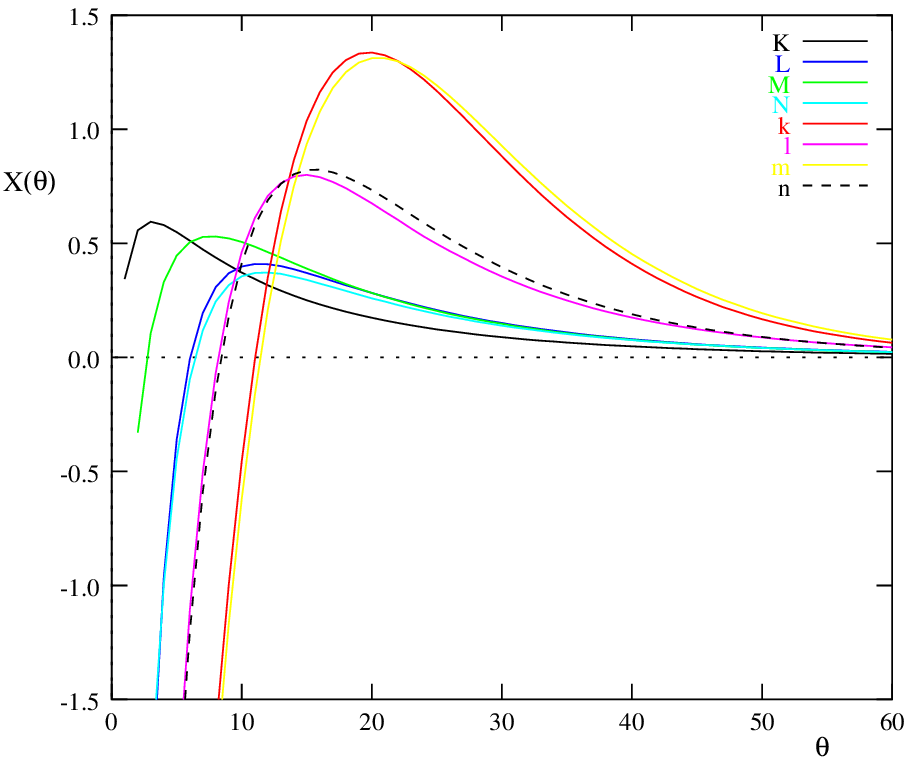,width=8.6cm,height=9cm}}
\end{picture}
\caption{%
$X(\theta)$ as a function of the threshold $\theta$ (in percent),
for all billiards shown in Fig.\ref{fig:pot3d}.}
\label{fig:eul95}
\end{figure}

\begin{figure}[t!!]
\setlength{\unitlength}{1cm}
\begin{picture}(17.4,9)
\put(0,0.2){\epsfig{file=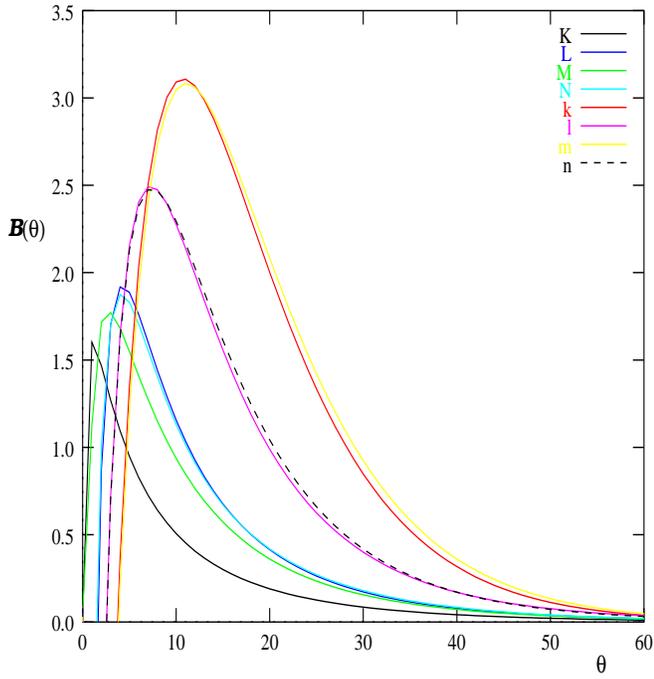,width=8.6cm,height=9cm}}
\end{picture}
\caption{%
$\mathbf {B} (\theta )$ as a function of the threshold $\theta$ (in percent),
for all billiards shown in Fig.\ref{fig:pot3d}.}
\label{fig:curv95}
\end{figure}

In order to look for a quantum signature of chaos in 3D
billiards (see Fig.\ref{fig:pot3d}) we apply an
analogous approach to the one described above for the 2D billiards.
In three dimensions both the mean breadth $B$ and the Euler characteristic
$\chi$ may be used to distinguish integrable and non-integrable billiards.
From Figs.\ref{fig:eul95} and \ref{fig:curv95} it is clear that
the $X(\theta )$ and $\mathbf B(\theta)$ curves form two groups depending on whether
the billiards are integrable or not. As was the case for the 2D billiards,
we can again use the maximum values of these curves to classify the
billiards according to their classical (non) integrability, as shown in Tables 2 and 3.
For classically integrable systems $\mu(B)$ lies
in the range [1.60,192] whereas for non-integrable systems $\mu (B)$ lies in the range
[2.47,3.10], a significant difference. For the Euler characteristic,
these ranges are [0.37-0.59] for integrable systems and [0.80,1.34] for
non-integrable ones. Here the separation is less in magnitude but still 
present.
\begin{table}[t]
\caption{$\mu (\chi)$ and the corresponding $\theta$ for all billiards shown in Fig.\ref{fig:pot3d}.}
\label{table:3}
\begin{center}
\begin{tabular}[h]{c|c|c}
 Billiard & $\theta$ & $\mu(\chi)$ \\
\hline
K & 3  & 0.59 \\
L & 11 & 0.41 \\
M & 8  & 0.53 \\
N & 12 & 0.37 \\
\hline
k & 20 & 1.34 \\
l & 15 & 0.80 \\
m & 21 & 1.31 \\
n & 16 & 0.82 \\
\end{tabular}
\end{center}
\end{table}
\begin{table}[t]
\caption{$\mu (B)$ and the corresponding $\theta$ for all billiards shown in Fig.\ref{fig:pot3d}.}
\label{table:2}
\begin{center}
\begin{tabular}[h]{c|c|c}
 Billiard & $\theta$ & $\mu(B)$ \\
\hline
K & 1  & 1.60 \\
L & 4  & 1.92 \\
M & 3  & 1.77 \\
N & 4  & 1.87 \\
\hline
k & 11 & 3.10 \\
l & 7  & 2.49 \\
m & 11 & 3.08 \\
n & 7  & 2.47 \\
\end{tabular}
\end{center}
\end{table}

\section{Transition to chaos}

\begin{figure}[t]
\setlength{\unitlength}{1cm}
\begin{picture}(8.6,5.7)
\put(0,0.2){\epsfig{file=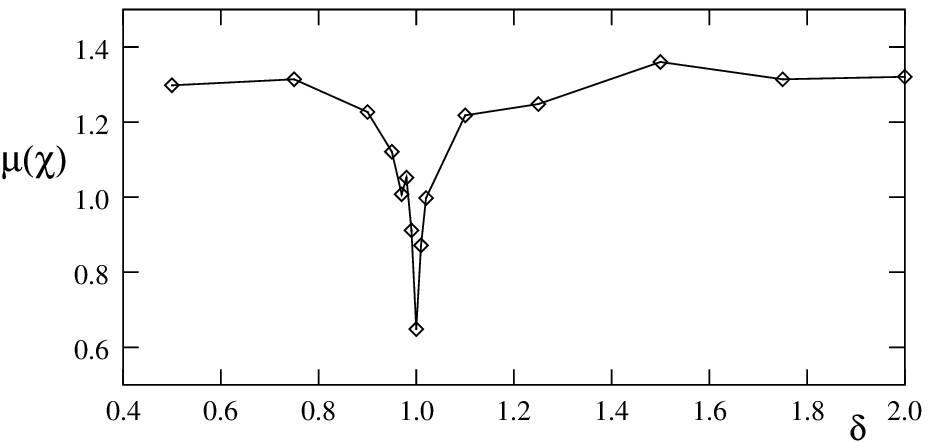,width=8.6cm,height=5.5cm}}
\end{picture}
\caption{%
$\mu(\chi)$ as function of the shape parameter $\delta$ for 
the lemon-shaped billiard. For $\delta=1$ the billiard is integrable.} 
\label{fig:fig9}
\end{figure}

An interesting case to confirm that $\mu(\chi)$ can be used to detect
the transition from integrable to chaotic behavior is the lemon-shaped
billiard \cite{Lopac}.
Two instances of the lemon-shaped billiard are shown in Fig.\ref{fig:fig4}.
The boundaries in the $x-y$ plane are defined by
$y(x)=\pm (1-|x|^{\delta})$, $x\in [-1,1]$.
The integrability of the
lemon-shaped billiard depends on the shape parameter $\delta$:
For $\delta =1$ and $\delta =\infty$ the billiard is integrable and otherwise it is not.
If $\delta$ slightly deviates from one,
the billiard is classically partially chaotic (periodic orbits and chaotic 
trajectories coexist).
Fig.\ref{fig:fig9} shows $\mu (\chi)$ for the lemon-shaped
billiard, for the shape parameter $\delta$ varying between 0.5 and 2.0.
The sharp peak centered around $\delta=1$
shows that $\mu (\chi )$ is very sensitive for the presence of
classically unstable trajectories. If there are any unstable trajectories in 
the classical system, the value of $\mu(\chi)$
for the quantum mechanical equivalent can be classified as chaotic. 
Thus, the curve in Fig. \ref{fig:fig9} clearly shows the 
transition from chaotic to integrable behavior and vice versa.

\section{Conclusion}
We have described a method, based on concepts of integral geometry,
to analyze the time evolution of the probability distribution of a
quantum particle moving in two and three-dimensional billiards.
We have demonstrated that the time-averaged Euler characteristic can be used
to classify billiards as integrable or not. For three-dimensional billiards
also the time-averaged mean breadth (or integral mean curvature) of the
probability density can be employed for this purpose.
For lemon-shaped billiards we have shown that the transition from integrable
to chaotic behavior can be determined from the dependence of the time-averaged
Euler characteristic on the parameter that controls the shape of the billiard.
We believe it may also be of interest to apply the method described in this
paper to the time-independent solutions.

\section{Acknowledgements}
We would like to thank the Dutch `Stichting Nationale Computer Faciliteiten' 
(NCF) for their support and A. Lande for his critical reading of the 
manuscript.

\end{document}